\documentclass[sigconf,natbib=true]{acmart}

\usepackage[utf8]{inputenc}
\usepackage{booktabs,multirow}
\usepackage{enumitem}
\usepackage{adjustbox}
\usepackage{color}
\usepackage{rotating}
\usepackage{amsmath}
\usepackage{pifont}
\usepackage{array}
\usepackage{bbm}
\usepackage[linesnumbered,algoruled,boxed,lined]{algorithm2e}

\newcommand{\ie}{{\it i.e.}}
\newcommand{\eg}{{\it e.g.}}

\newcommand{\ul}{\underline}{}

\newcommand{\argmax}{\operatornamewithlimits{argmax}}

\AtBeginDocument{%
  }

\copyrightyear{2024}
\acmYear{2024}
\setcopyright{rightsretained}
\acmConference[SIGIR '24]{Proceedings of the 47th International ACM SIGIR
Conference on Research and Development in Information Retrieval}{July 14--18,
2024}{Washington, DC, USA}
\acmBooktitle{Proceedings of the 47th International ACM SIGIR Conference on
Research and Development in Information Retrieval (SIGIR '24), July 14--18,
2024, Washington, DC, USA}
\acmDOI{10.1145/3626772.3657928}
\acmISBN{979-8-4007-0431-4/24/07}

\begin{document}

\title{Multi-intent-aware Session-based Recommendation}

\author{Minjin Choi}
\affiliation{
  \institution{Sungkyunkwan University}
  \country{Republic of Korea}}
\email{zxcvxd@skku.edu}

\author{Hye-young Kim}
\affiliation{
  \institution{Sungkyunkwan University}
  \country{Republic of Korea}}
\email{khyaa3966@skku.edu}

\author{Hyunsouk Cho}
\affiliation{
  \institution{Ajou University}
  \country{Republic of Korea}}
\email{hyunsouk@ajou.ac.kr}

\author{Jongwuk Lee}\authornote{Corresponding author}
\affiliation{
  \institution{Sungkyunkwan University}
  \country{Republic of Korea}}
\email{jongwuklee@skku.edu}

\begin{abstract}

Session-based recommendation (SBR) aims to predict the following item a user will interact with during an ongoing session. Most existing SBR models focus on designing sophisticated neural-based encoders to learn a session representation, capturing the relationship among session items. However, they tend to focus on the last item, neglecting diverse user intents that may exist within a session. This limitation leads to significant performance drops, especially for longer sessions. To address this issue, we propose a novel SBR model, called \emph{Multi-intent-aware Session-based Recommendation Model (\textbf{MiaSRec})}. It adopts frequency embedding vectors indicating the item frequency in session to enhance the information about repeated items. MiaSRec represents various user intents by deriving multiple session representations centered on each item and dynamically selecting the important ones. Extensive experimental results show that MiaSRec outperforms existing state-of-the-art SBR models on six datasets, particularly those with longer average session length, achieving up to 6.27\% and 24.56\% gains for MRR@20 and Recall@20. Our code is available at \url{https://github.com/jin530/MiaSRec}.
\end{abstract}



\maketitle

\section{Introduction}

\begin{figure}
\includegraphics[width=0.9\linewidth]{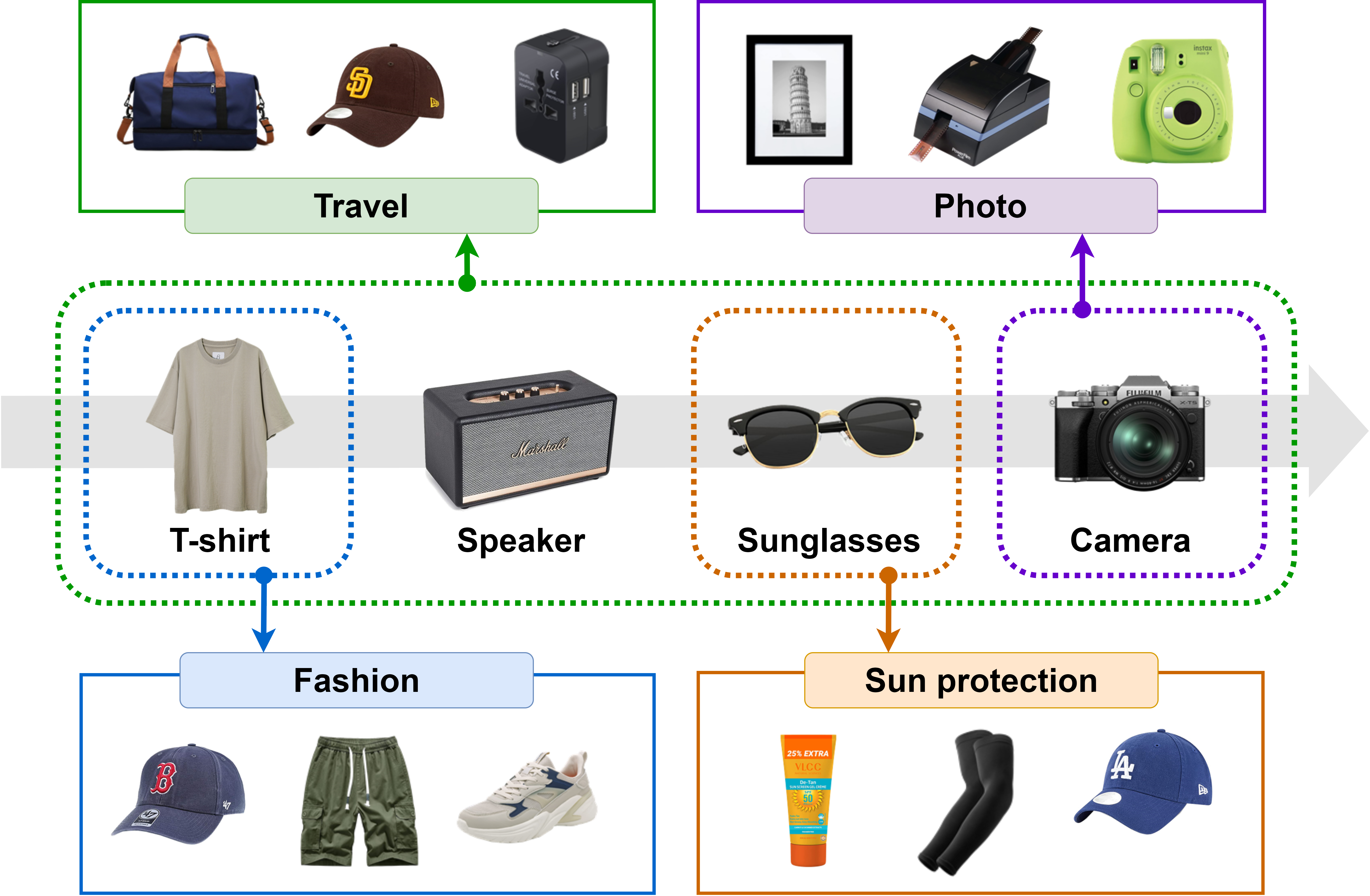}
\caption{A session example with multiple user intents, such as \emph{travel}, \emph{fashion}, \emph{sun protection}, and \emph{photo}. Dotted rectangles represent items related to each user intent, and solid rectangles represent recommendations for each user intent.}\label{fig:motivation}
\vskip -0.15in  
\end{figure}

Session-based recommendation (SBR)~\cite{JannachLL17SRsurvey, WangCW19SRsurvey} aims to learn hidden user preferences in a session and provide personalized items for each user. A session refers to a sequence of user-item interactions over time, \eg, consecutive clicks on multiple products during a transaction. It is particularly effective for anonymous or first-time users in web applications like e-commerce and streaming services, \eg, Amazon, YouTube, Netflix, and Spotify~\cite{LindenSY03Amazon, CovingtonAS16, Gomez-UribeH16Netflix, ChenLSZ18Spotify}. SBR inherently suffers from extreme data sparsity because it only deals with user actions during an ongoing session, making it challenging to capture dynamic and intricate item correlations.

Existing SBR models~\cite{HidasiKBT15GRU4Rec, LiRCRLM17NARM, HidasiK18GRU4Rec+, WuT0WXT19SRGNN, PanCCCR20SGNNHN, WangWCLMQ20GCEGNN, KangM18SASRec, YuanSSWZ21DSAN} have primarily focused on extracting a single representation from a session to capture and express user preferences. They mainly aimed to model a session consisting of multiple items using various neural-based session encoders, including recurrent neural networks (RNNs)~\cite{HidasiKBT15GRU4Rec, LiRCRLM17NARM, HidasiK18GRU4Rec+}, graph neural networks (GNNs)~\cite{WuT0WXT19SRGNN, PanCCCR20SGNNHN, WangWCLMQ20GCEGNN}, or Transformers~\cite{KangM18SASRec, YuanSSWZ21DSAN}. However, despite their advanced encoder designs, modeling only a single representation cannot express multiple user intents.

Figure~\ref{fig:motivation} illustrates the importance of using multiple user intents. While the user may be interested in \emph{photo} when focusing on the last item ``camera'', looking at the entire session suggests that the user will click items about \emph{travel}. Considering the other items in the session, \emph{fashion} or \emph{sun protection} also align with different user intents. In this scenario, it is more appropriate to recommend a top-$N$ item list with appropriate multiple intents, \eg, (``travel bag'', ``sneakers'', and ``photo frame''). On the other hand, in some sessions, not all items are important. For example, in Figure~\ref{fig:motivation}, ``speaker'' is less relevant to the other items for capturing user intents.

Recently, some SBR models, such as MSGIFSR~\cite{Guo0SZWBZ22MSGIFSR} and Atten-Mixer~\cite{ZhangGLXKZXWK23Atten-Mixer}, have attempted to capture multiple user intents, focusing primarily on the last few consecutive items to represent diverse user intents. However, they cannot accurately capture user intent if the last item is less important or noisy. Besides, some studies~\cite{Wang0WSOC19MCPRN, TanZYLZYH21SINE, ChenZZXX21ComiRec, ZhangYYLFZC022Re4} have attempted to identify multiple user interests over a long user-item history. They employ a fixed number of user interests and extract the same number of interests for all users. It may miss some interests or include unnecessary ones since the number of user interests varies by user. We claim the challenges for modeling multiple user intents in the session: (i) \emph{how to fully capture multiple user intents inherent in each session} and (ii) \emph{how to filter out unimportant ones among multiple intents}. 

\begin{figure}
\includegraphics[width=0.85\linewidth]{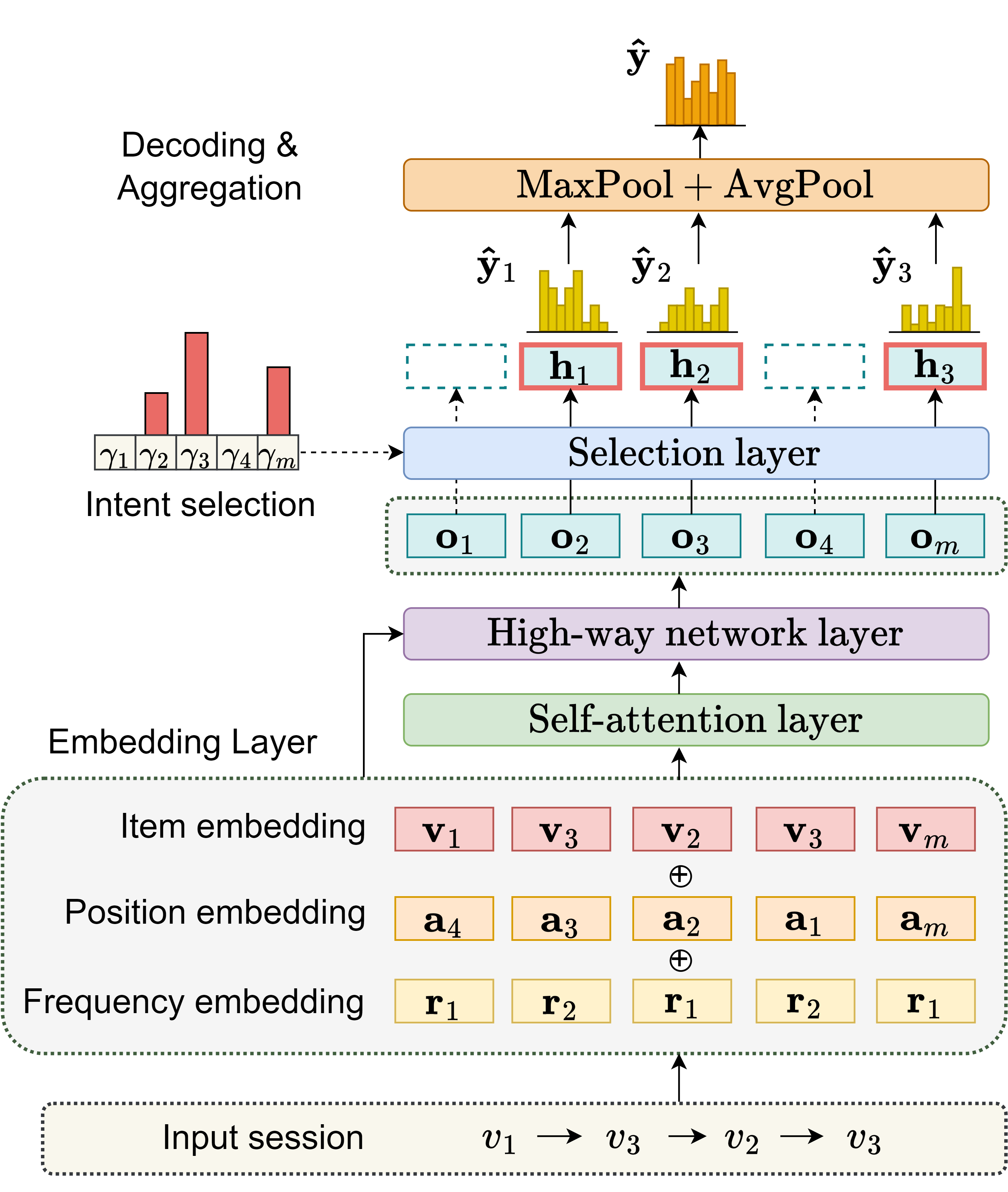}
\caption{The model architecture of MiaSRec.}\label{fig:architecture}
\vskip -0.2in
\end{figure}

To address these issues, we propose a novel SBR model, called \emph{Multi-intent-aware Session-based Recommendation Model (\textbf{MiaSRec})}, as shown in Figure~\ref{fig:architecture}. First, MiaSRec encodes the session items with position and frequency embeddings, reflecting sequential information and repeat patterns. It then employs a self-attention mechanism and a high-way network to derive different user intents in the session. Then, it adaptively extracts diverse user intents in the session. Lastly, MiaSRec decodes multiple session representations into item distributions and aggregates them using pooling functions. Despite its simplicity, extensive experiments demonstrate that MiaSRec outperforms existing SBR models on six benchmark datasets. Notably, MiaSRec achieves significant gains for longer sessions ($\ge10$) with multiple user intents, up to 13.51\% in Recall@20.
\section{Proposed Model}\label{sec:framework}

\subsection{Session-based Recommendation}\label{sec:problem_statement}
Let $\mathcal{V} = \{v_1, \dots, v_{n}\}$ denote a set of $n$ unique items, \eg, products and songs. An arbitrary session $s = (v_{t_1}, \dots, v_{t_{|s|}})$ represents a sequence of $|s|$ items that a user interacts with, \eg, clicks, views, and purchases. Here, $t_i$ indicates the index of the $i$-th item in the session. Given a session, the goal of SBR is to predict the next item $v_{t_{|s|+1}}$ that the user is likely to consume. The SBR model takes a session as an input and returns a top-$N$ item list to recommend. 

\subsection{Model Architecture}\label{sec:encoder}

In this section, we present a novel SBR model called MiaSRec, which aims to address (i) how to represent multiple user intents in the session and (ii) how to prune out unnecessary user intents. 

\subsubsection{Embedding Layer}\label{sec:embedding_layer}

We first embed each session item $v_{t_i}$ into item embedding vector $\mathbf{v}_{t_i} \in \mathbb{R}^{d}$, and generate the mean item embedding $\mathbf{v}_{m} = \frac{1}{|s|}\sum_{i=1}^{|s|}{\mathbf{v}_{t_i}}$, capturing the global session information. To better capture the importance of each item in a session, we incorporate the absolute position embedding vector $\mathbf{a}_i \in \mathbb{R}^{d}$ to distinguish sequential order and the frequency embedding vector $\mathbf{r}_{f_i} \in \mathbb{R}^{d}$ to express the importance of repeated items in a session. Here, $f_i$ denotes the frequency of the $i$-th item in a session. In Figure~\ref{fig:architecture}, given session $s = (v_1, v_3, v_2, v_3)$, the frequency is $(1, 2, 1, 2)$. Finally, the item, position, and frequency embedding vectors are combined as the model input $x_i$.
\begin{equation}
\label{eq:input}
\mathbf{x}_{i} = \mathbf{v}_{t_i} + \mathbf{a}_{i} + \mathbf{r}_{f_i}  \ \text{for} \ i \in \{1, \dots, |s|, m\} .
\end{equation}

\noindent
Note that we sort session items in reverse order as in ~\cite{WangWCLMQ20GCEGNN}, so the last item $v_{t_{|s|}}$ always corresponds to the first positional embedding $\mathbf{a}_1$. And, we randomly initialize the learnable parameter of position and frequency embeddings.

\subsubsection{Multi-intent Representation}
We employ the self-attention mechanism~\cite{VaswaniSPUJGKP17Transformer} to capture the complex relationships among session items. Using the bi-directional self-attention layer $\text{Self-attention}(\cdot)$, we generate multiple contextualized representations of the session information associated with each item as follows:
\begin{equation}
\label{eq:transformer_output}
\mathbf{c}_{1}, \dots, \mathbf{c}_{|s|}, \mathbf{c}_{m} = \text{Self-attention}([\mathbf{x}_{1}, \dots, \mathbf{x}_{|s|}, \mathbf{x}_{m}])).
\end{equation}


To ensure that multiple contextualized representations do not become similar and to better reflect different user intent, we leverage the high-way network~\cite{PanCCCR20SGNNHN} by emphasizing item embeddings. Specifically, we combine contextualized representation $\mathbf{c}_{i}$ and item embedding $\mathbf{v}_{i}$ to derive user intent representation $\mathbf{o}_{i} \in \mathbb{R}^{d}$ that takes into account both the overall information of the session and the information of each item.
\begin{equation}
\label{eq:highway}
\begin{split}
\mathbf{o}_{i} = \mathbf{g} \odot  \mathbf{v}_{i} + (1-\mathbf{g}) & \odot \mathbf{c}_{i} \ \text{for} \ i \in \{1, \dots, |s|, m\},  \\
 \text{where} \ \mathbf{g} =  \sigma( & \mathbf{W_{g}}[\mathbf{v}_{i};\mathbf{c}_{i}]^{\top}),
\end{split}
\end{equation}
\noindent
where $\mathbf{W_{g}} \in \mathbb{R}^{d \times 2d}$ is a learnable weight matrix, $\mathbf{g} \in \mathbb{R}^{d}$ is a gating vector, and $\sigma(\cdot)$ is the sigmoid function. 


\subsubsection{Intent Selection} We employ multiple session representations to fully exploit the potential of each session item. However, not all session items may be necessary, and some may be noisy. To extract essential user intents in a session, we calculate the importance of multiple representations and remove unimportant ones. We utilize a sparse transformation $\alpha$-entmax~\cite{PetersNM19entmax, YuanSSWZ21DSAN}, assigning a zero probability to unimportant representations. 
\begin{equation}
\alpha\text{-entmax}(\mathbf{z}) = \argmax_{ \mathbf{p} \in \Delta^{l}}{ \mathbf{p}^{\top} \mathbf{z}+H^{T}_{\alpha}(\mathbf{p})},
\end{equation}
where $H^{T}_{\alpha}( \mathbf{p})=\frac{1}{\alpha(\alpha-1)}\sum_{j}( \mathbf{p}_j - \mathbf{p}_{j}^{\alpha})$ if $\alpha \neq 1$, else $H^{T}_{1}(p)=-\sum_{j} \mathbf{p}_j \log \mathbf{p}_j$~\footnote{$\Delta^{l}:=\{\mathbf{p} \in \mathbb{R}^{l}: \mathbf{p} \ge 0, \|p\|_1=1$\} denotes the $l$-probability simplex.}. Depending on $\alpha$, 1-entmax and 2-entmax are equivalent to softmax and sparsemax~\cite{MartinsA16sparsemax}, respectively. A larger $\alpha$ value generates a more sparse probability distribution, and we set $\alpha$ as 1.5 empirically. We extract session representation $\mathbf{h}_{i} \in \mathbb{R}^{d}$ by masking unnecessary user intents.
\begin{equation}
\begin{split}
\label{eq:extractor}
\mathbf{\gamma} = \alpha & \text{-entmax}(\mathbf{w} \cdot [\mathbf{o}_{1}; \dots; \mathbf{o}_{|s|}; \mathbf{o}_{m}]^{\top}),  \\
\{ \mathbf{h}_{1}, \dots, \mathbf{h}_{k} \} & = \{ \mathbf{\gamma}_{i} \mathbf{o}_{i} | \mathbf{\gamma}_{i} > 0, i \in \{1, \dots, |s|, m\} \},
\end{split}
\end{equation}
where $\mathbf{w} \in \mathbb{R}^{d}$ is a learnable parameter and $\mathbf{\gamma} \in \mathbb{R}^{|s|+1}$ is the importance weight vector for each item. $k$ is the number of non-zero elements in $\gamma$. Unlike the previous multiple representation models~\cite{ZhangYYLFZC022Re4, ChenZZXX21ComiRec, Guo0SZWBZ22MSGIFSR, ZhangGLXKZXWK23Atten-Mixer} utilize a fixed number of representations regardless of the session, MiaSRec dynamically selects multiple session representations for a given session, up to the number of session items.

\subsubsection{Multi-intent Aggregation}
The multi-intent aggregation process of MiaSRec is divided into two parts: (i) decoding item distributions from multiple session representations and (ii) aggregating the distributions for the final recommendation. 

To decode each session representation into the item distribution, we employ cosine similarity with item embedding matrix, \ie, dot product with L2-normalization. For simplicity, we reuse the item embedding look-up table $\mathbf{V} \in \mathbb{R}^{n \times d}$, so the number of model parameters does not increase.
\begin{equation}
\label{eq:decoder}
\mathbf{\hat{y}}_{1}, \dots, \mathbf{\hat{y}}_{k} = [\mathbf{\Tilde{h}}_{1}\mathbf{\Tilde{V}}^{\top}, \dots, \mathbf{\Tilde{h}}_{k}\mathbf{\Tilde{V}}^{\top}],
\end{equation}
\noindent
where $\mathbf{\Tilde{h}}_{i}$ and $\mathbf{\Tilde{V}}$ are the normalized session vector and the normalized item embedding matrix, respectively. $\mathbf{\hat{y}}_{i} \in \mathbb{R}^{n}$ represents an item distribution decoded by the session vector $\mathbf{h}_{i}$.

To aggregate multiple item distributions, we adopt max-pooling and average-pooling functions. Max-pooling maintains the principal features for multiple intents, and average-pooling captures consistent intent over the session. 
\begin{equation}
\label{eq:maxpool}
\mathbf{\hat{y}} = \beta \text{MaxPool}(\mathbf{\hat{y}}_{1}, \dots, \mathbf{\hat{y}}_{k}) + (1-\beta)  \text{AvgPool}(\mathbf{\hat{y}}_{1}, \dots, \mathbf{\hat{y}}_{k}),
\end{equation}
\noindent
where $\mathbf{\hat{y}} \in \mathbb{R}^{n}$ is the final aggregated item distribution, and $\beta$ is a combination hyperparameter. $\text{MaxPool}(\cdot)$ and $\text{AvgPool}(\cdot)$ represent the max- and average-pooling functions that fetch the maximum (or average) value for each dimension from multiple vectors.

Lastly, we formulate a cross-entropy loss function for training.
\begin{equation}
\label{eq:recommendation_loss}
    L(\mathbf{y},\mathbf{\hat{y}}) = - \sum^{n}_{j=1} \mathbf{y}(j)\log(\frac{\text{exp}(\mathbf{\hat{y}}(j) / \tau)}{\sum_i{\text{exp}(\mathbf{\hat{y}}(i) / \tau)}}),
\end{equation}
\noindent
where $\mathbf{y} \in \mathbb{R}^{n}$ is a one-hot vector of the target item, \ie, $\mathbf{y}(j)=1$ if the $j$-th item is the target item; otherwise, $\mathbf{y}(j)=0$. Here, $\tau$ is the hyperparameter to control the temperature~\cite{HintonVD15distillation} for better convergence~\cite{GuptaGMVS19NISER}.

\section{Experiments}\label{sec:exp_all}
\subsection{Experimental Setup}\label{sec:setup}

\begin{table}[] \small
\caption{Statistics of the various benchmark datasets. AvgLen indicates the average length of entire sessions.}\label{tab:dataset}
\begin{tabular}{c|cccc}
\hline
Dataset      & \# Interacts & \# Sessions & \# Items & AvgLen \\ \hline
Diginetica   & 786,582         & 204,532     & 42,862   & 4.12        \\
Retailrocket & 871,637         & 321,032     & 51,428   & 6.40        \\
Yoochoose    & 1,434,349       & 470,477     & 19,690   & 4.64        \\
Tmall        & 427,797         & 66,909      & 37,367   & 10.62       \\
Dressipi     & 4,305,641       & 943,658     & 18,059   & 6.47        \\
LastFM       & 3,510,163       & 325,543     & 38,616   & 8.16        \\ \hline
\end{tabular}
\end{table}

\begin{table*} \small
\centering
\setlength{\tabcolsep}{3pt} 
\caption{Performance comparison for MiaSRec and baseline models. Imp. indicates how much better MiaSRec is than the best baseline model. The best model is marked {\color[HTML]{FF0000} \textbf{bold}} and the second best model is {\color[HTML]{0000FF} \ul{underlined}}. Significant differences ($\rho < 0.01$) between the best baseline model and MiaSRec are reported with $\dagger$.}\label{tab:exp_all}
\begin{tabular}{c|c|ccccccc|ccccc|c|c}
\hline
Dataset & Metric & SASRec & SR-GNN & NISER+ & SGNN-HN & DSAN & LESSR & CORE & SINE & ComiRec & Re4 & A-mixer & MSGIFSR & MiaSRec & Imp. \\ \hline
 & R@20 & 49.86 & 48.01 & 51.11 & 50.60 & 52.06 & 48.70 & 52.89 & 46.45 & 51.22 & 51.59 & 49.84 & {\color[HTML]{0000FF} \ul{53.20}} & {\color[HTML]{FF0000} \textbf{53.54}} & 0.65\% \\
\multirow{-2}{*}{Diginetica} & M@20 & 17.20 & 16.60 & 18.21 & 17.28 & 18.25 & 16.96 & {\color[HTML]{0000FF} \ul{18.53}} & 16.10 & 18.35 & 18.47 & 17.07 & 18.37 & {\color[HTML]{FF0000} \textbf{19.47}}$^{\dagger}$ & 5.04\% \\ \hline
 & R@20 & 59.70 & 58.01 & 60.70 & 57.43 & 61.13 & 56.56 & 61.77 & 55.11 & 61.56 & 61.65 & 59.49 & {\color[HTML]{0000FF} \ul{63.04}} & {\color[HTML]{FF0000} \textbf{63.37}} & 0.26\% \\
\multirow{-2}{*}{Retailrocket} & M@20 & 35.71 & 36.01 & 38.18 & 35.39 & {\color[HTML]{0000FF} \ul{38.68}} & 36.82 & 38.49 & 34.15 & 38.16 & 38.10 & 36.25 & 38.42 & {\color[HTML]{FF0000} \textbf{39.23}}$^{\dagger}$ & 1.41\% \\ \hline

 & R@20 & 63.64 & 62.28 & 63.50 & 61.60 & 63.73 & 62.78 & 64.64 & 57.50 & 63.13 & 62.67 & 63.73 & {\color[HTML]{0000FF} \ul{65.20}} & {\color[HTML]{FF0000} \textbf{65.37}} & 0.26\% \\
\multirow{-2}{*}{Yoochoose} & M@20 & 28.66 & 28.36 & 29.06 & 27.97 & 29.23 & 28.84 & 28.25 & 25.07 & 28.29 & 28.00 & 29.32 & {\color[HTML]{0000FF} \ul{30.02}} & {\color[HTML]{FF0000} \textbf{30.74}}$^{\dagger}$ & 2.39\% \\ \hline 

& R@20 & 35.80 & 33.47 & 40.39 & 39.71 & 42.82 & 32.59 & {\color[HTML]{0000FF} \ul{44.91}} & 35.66 & 42.40 & 41.56 & 38.76 & 35.39 & {\color[HTML]{FF0000} \textbf{55.94}}$^{\dagger}$ & 24.56\% \\
\multirow{-2}{*}{Tmall} & M@20 & 25.08 & 24.75 & 29.48 & 24.16 & 30.85 & 24.19 & {\color[HTML]{0000FF} \ul{31.59}} & 22.41 & 28.43 & 28.56 & 28.52 & 22.19 & {\color[HTML]{FF0000} \textbf{33.57}}$^{\dagger}$ & 6.27\% \\ \hline

 & R@20 & 37.18 & 36.10 & 38.19 & 38.35 & 37.77 & 37.71 & 38.14 & 38.18 & {\color[HTML]{0000FF} \ul{39.60}} & 39.15 & 37.75 & 38.43 & {\color[HTML]{FF0000} \textbf{42.26}}$^{\dagger}$ & 6.73\% \\
\multirow{-2}{*}{Dressipi} & M@20 & 14.31 & 14.51 & 15.34 & 15.05 & 15.13 & 14.73 & 15.54 & 15.46 & {\color[HTML]{0000FF} \ul{16.07}} & 15.92 & 15.24 & 15.90 & {\color[HTML]{FF0000} \textbf{16.70}}$^{\dagger}$ & 3.92\% \\ \hline
 & R@20 & 20.53 & 21.80 & 22.50 & 22.72 & 22.47 & 22.31 & 22.75 & 22.17 & 22.13 & {\color[HTML]{0000FF} \ul{23.02}} & 22.93 & 22.73 & {\color[HTML]{FF0000} \textbf{25.85}}$^{\dagger}$ & 12.32\% \\
\multirow{-2}{*}{LastFM} & M@20 & 6.22 & 8.70 & 8.79 & 7.66 & 7.93 & {\color[HTML]{0000FF} \ul{8.80}} & 7.83 & 7.57 & 7.83 & 8.50 & 8.74 & 8.20 & {\color[HTML]{FF0000} \textbf{9.95}}$^{\dagger}$ & 13.06\% \\ \hline
\end{tabular}
\end{table*}

\textbf{Datasets.} We conduct extensive experiments on six real-world datasets collected from e-commerce and music streaming services: Diginetica, Retailrocket, Yoochoose, Tmall~\footnote{Since it has been widely used in previous studies~\cite{hou2022core, 0013YYSC21COTREC, HanZCLSL22MGIR}, we adopt Tmall even though it consists of timestamps in units of days, not in minutes or seconds.}, Dressipi, and LastFM. For data pre-processing, we follow the conventional procedure~\cite{LudewigJ18Evaluation, LudewigMLJ19bEmpiricalAnalysis, LiRCRLM17NARM, WuT0WXT19SRGNN}. We discard the sessions with a single item and the items that occur less than five times in entire sessions. We split training, validation, and test sets chronologically as the 8:1:1 ratio. Table~\ref{tab:dataset} summarizes detailed statistics on all benchmark datasets.

\noindent
\textbf{Baselines.} We compare MiaSRec with the following SBR models: \textbf{SASRec}~\cite{KangM18SASRec}, \textbf{SR-GNN}~\cite{WuT0WXT19SRGNN}, \textbf{NISER+}~\cite{GuptaGMVS19NISER}, \textbf{SGNN-HN}~\cite{PanCCCR20SGNNHN}, \textbf{DSAN}~\cite{YuanSSWZ21DSAN}, \textbf{LESSR}~\cite{ChenW20LESSR}, \textbf{CORE}~\cite{hou2022core}. We also compare with the subsequent multiple representations models. \textbf{SINE}~\cite{TanZYLZYH21SINE}, \textbf{ComiRec}~\cite{ChenZZXX21ComiRec}, \textbf{Re4}~\cite{ZhangYYLFZC022Re4}, \textbf{Atten-mixer (A-mixer)}~\cite{ZhangGLXKZXWK23Atten-Mixer}, \textbf{MSGIFSR}~\cite{Guo0SZWBZ22MSGIFSR}. We do not consider SBR models which use additional information, \eg, temporal information~\cite{gng-ode, TMIGNN} or content-based features~\cite{HidasiQKT16p-RNN, ZhuCLYYX20KA-MemNN, Chen0ZSW22Satori, LiWZMCZ22ISRec}.

\noindent
\textbf{Evaluation protocol and metrics.} As the common protocol to evaluate SBR models~\cite{LiRCRLM17NARM, WuT0WXT19SRGNN}, we adopt the \emph{iterative revealing scheme}, which iteratively exposes an item from a session to the model. We adopt Recall (R@20) and Mean Reciprocal Rank (M@20) to quantify the prediction accuracy of the next single item. All experimental results are averaged over three runs with different seeds, and we conduct the significance test using a paired t-test. 

\noindent
\textbf{Implementation details.} For reproducibility, we implement MiaSRec and the other baseline models on an open-source recommendation system library RecBole\footnote{https://github.com/RUCAIBox/RecBole}~\cite{ZhaoMHLCPLLWTMF21RecBole, ZhaoHPYZLZBTSCX22RecBole2.0}. We optimize all baselines using Adam optimizer~\cite{KingmaB14Adam} with a learning rate of 0.001. We set the embedding dimension to 100 and the max session length to 50. We stop the training if the validation MRR@20 decreases for three consecutive epochs~\footnote{We report the performance on the test set using the models that show the highest performance on the validation set.}. For all methods, we set the batch size to 1024. For MiaSRec, we set $\alpha$ as 1.5 for $\alpha$-entmax and tune the temperature $\tau$ among \{0.01, 0.05, 0.07, 0.1, 0.5, 1\}, dropout ratio $\delta$ among \{0, 0.1, 0.2, 0.3, 0.4, 0.5\}. We search $\beta$ from 0 to 1 in 0.1 increments. We follow the original papers' settings for other hyperparameters of baseline models, but if not available, we thoroughly tune them. 







\subsection{Experimental Results}\label{sec:results}

\textbf{Overall comparison.} Table~\ref{tab:exp_all} reports the performance comparison between MiaSRec and other baseline models. (i) MiaSRec shows the best performance on all datasets. Note that MiaSRec demonstrates performance improvements of up to 24.56\% for R@20 compared to the best baseline. In particular, MiaSRec shows substantial improvements in Recall on datasets with longer average session lengths (\eg, Tmall and LastFM). (ii) Multiple representation models, such as MSGIFSR~\cite{Guo0SZWBZ22MSGIFSR}, ComiRec~\cite{ChenZZXX21ComiRec} and Re4~\cite{ZhangYYLFZC022Re4}, tend to show higher accuracy than single representation models. This implies that various intents can exist in session, and it is necessary to capture them. 
\begin{figure}[t]
\centering
\begin{tabular}{c}
\includegraphics[width=0.44\textwidth]{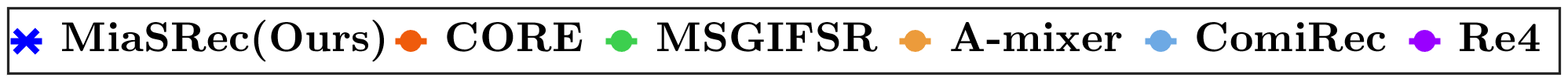} \\
\end{tabular}
\begin{tabular}{cc}
\includegraphics[width=0.23\textwidth]{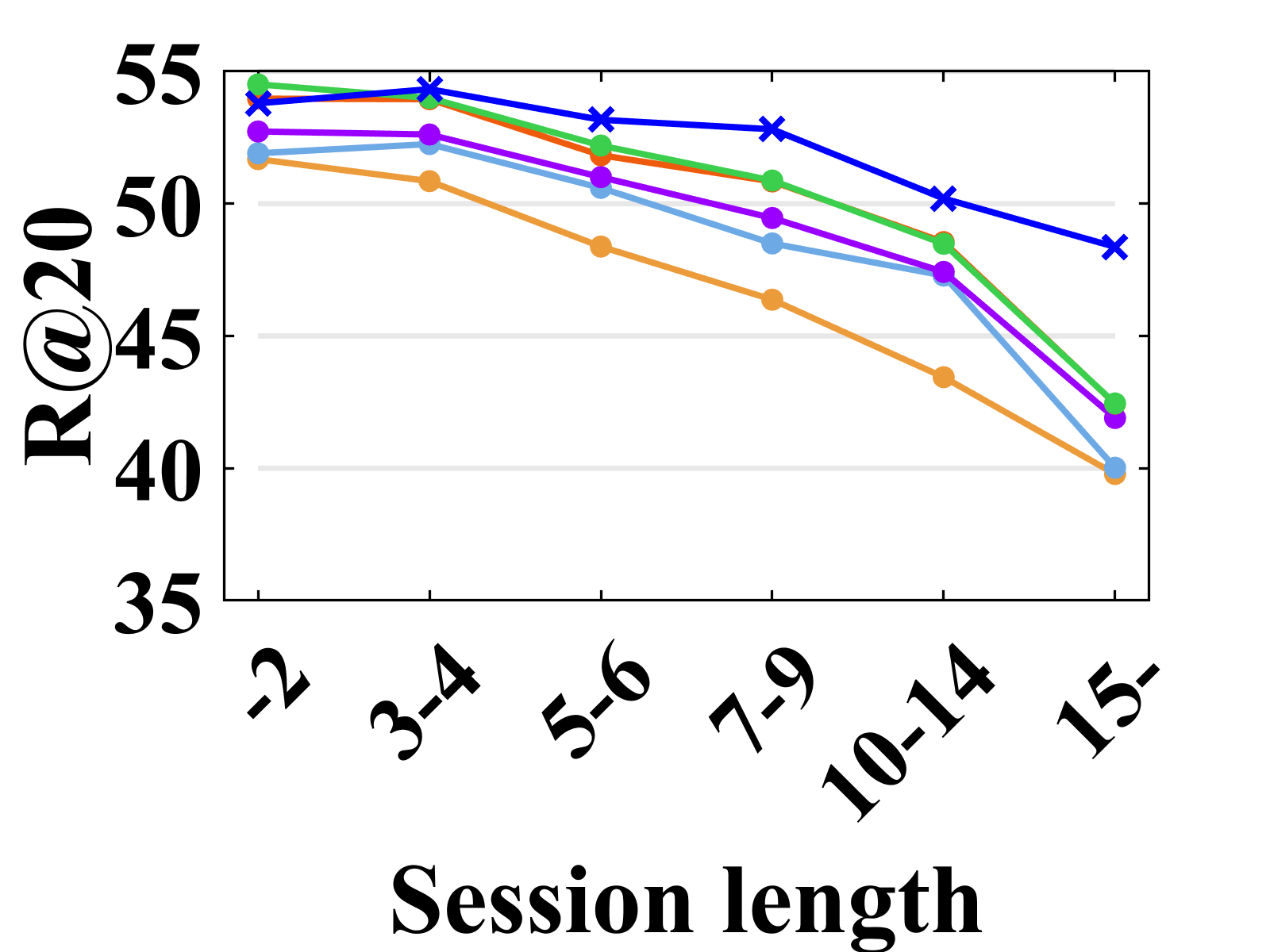}  &
\includegraphics[width=0.23\textwidth]{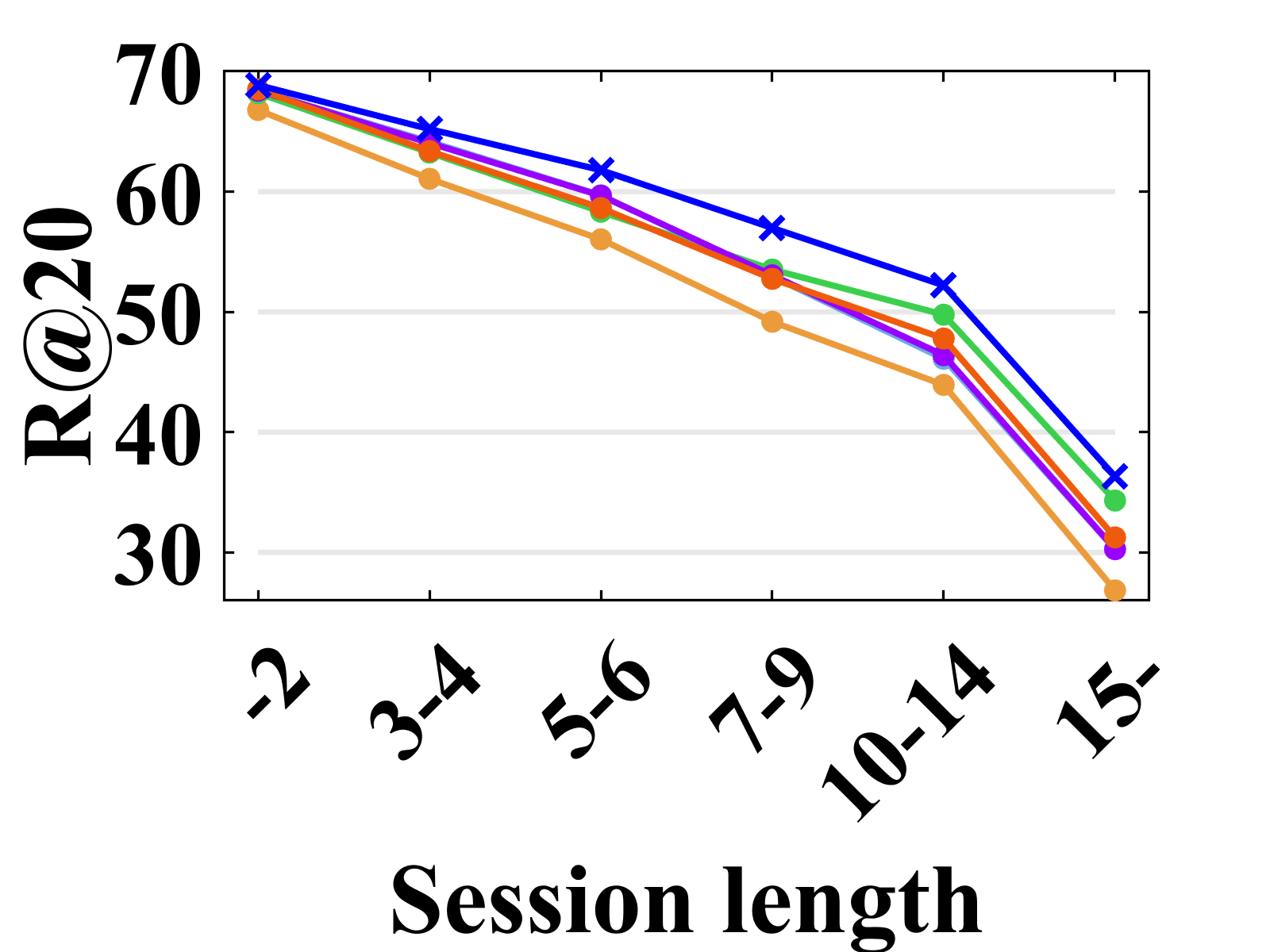} \\
(a) Diginetica & (b) Retailrocket \\
\end{tabular}
\caption{Performance comparison of SBR models over varying session lengths. Sessions are divided into six groups depending on session length.}\label{fig:exp_session_length}
\vspace{-3mm}
\end{figure}

%

\vspace{1mm}
\noindent
\textbf{Effect of session length.} Figure~\ref{fig:exp_session_length} illustrates the accuracy of SBR models as the session length varies. (i) The accuracy of all models decreases as the session length increases since user intent can vary. For example, in Diginetica, CORE~\cite{hou2022core} shows 13.2\% performance drop for long sessions ($|s|$$\ge$$10$) compared to short sessions ($|s|$$<$$5$). (ii) MiaSRec shows the highest performance in most cases, regardless of session length, and a comparatively modest performance drop, which indicates that MiaSRec effectively captures various user intents. Particularly, for long sessions ($|s|$$\ge$$10$), it shows significant improvements over CORE, \eg, 6.03\% for R@20 in Diginetica. 


\begin{table}[] \small
\centering
\renewcommand{\arraystretch}{0.95}
\caption{Ablation study for MiaSRec. ``PE'' and ``FE'' mean position and frequency embedding. ``mean'' indicates only using the mean vector ($\textbf{o}_{\text{m}}$), and ``last $k$'' indicates selecting the last $k$ representations as session representations. }\label{tab:exp_ablation}
\begin{adjustbox}{width=\columnwidth,center}
\begin{tabular}{c|cc|cc|cc}
\hline
\multirow{2}{*}{Model} & \multicolumn{2}{c|}{Diginetica} & \multicolumn{2}{c|}{Retailrocket} & \multicolumn{2}{c}{Yoochoose} \\
 & R@20 & M@20 & R@20 & M@20 & R@20 & M@20 \\ \hline
\textbf{MiaSRec} & \textbf{53.54} & \textbf{19.47} & \textbf{63.37} & \textbf{39.23} & \textbf{65.37} & \textbf{30.74} \\
\hline
\multicolumn{7}{c}{\textit{Variants for embedding layers}} \\ 
\hline
w/o PE ($\mathbf{a}_i$) & 51.36 & 18.15 & 61.06 & 37.51 & 61.13 & 26.51 \\
w/o FE ($\mathbf{r}_{f_i}$) & 53.48 & 19.23 & 63.28 & 38.92 & 65.15 & 29.91 \\
\hline
\multicolumn{7}{c}{\textit{Variants for intent selection}} \\ 
\hline
mean ($\mathbf{o}_{{m}}$) &  52.73 & 18.66 & 61.70 & 38.25 & 64.71 & 29.84 \\
last $1$ ($\mathbf{o}_{{|s|}}$) &  52.34 & 18.37 & 61.90 & 37.79 & 64.10 & 30.05 \\
last $3$ ($\mathbf{o}_{{|s|-2:|s|}}$) &  53.08 & 19.20 & 62.07 & 38.68 & 64.77 & 30.34 \\
last $5$ ($\mathbf{o}_{{|s|-4:|s|}}$) &  53.38 & 19.29 & 63.01 & 38.90 & 65.13 & 30.51 \\
\hline
\end{tabular}
\end{adjustbox}
\end{table}


\vspace{1mm}
\noindent
\textbf{Ablation study.} Table~\ref{tab:exp_ablation} shows the ablation study of MiaSRec for additional embeddings and multi-intent selection. (i) Both frequency and position embeddings have a significant impact on performance. This indicates that reflecting the importance of each item through sequential and occurrence information is effective in improving performance. (ii) It is always better to use multiple representations than a single representation. In particular, MiaSRec shows up to 2.71\% improvements in R@20 compared to single representation variants using mean vector ($\mathbf{o}_{m}$) and last item vector ($\mathbf{o}_{|s|}$). (iii) The intent selection method of MiaSRec is more effective than heuristic multiple representation variants. It outperforms other methods that adopt multiple representations of the last few item vectors, suggesting the importance of extracting the representation dynamically over the session.

\section{Conclusion}\label{sec:conclusion}

This paper proposed a novel SBR model, \textbf{MiaSRec}, which exploits various user intents in a session. Unlike previous SBR models that only use a single session representation, MiaSRec fully captures a variety of intents utilizing each session item using multiple representations and dynamically selects more important ones using the intent selection layer. It then effectively decodes and aggregates the multiple representations and provides recommendations that reflect the various intents. Extensive experiments showed that MiaSRec outperformed twelve baseline models on six benchmark datasets. 

\section*{Acknowledgement}
This work was supported by Institute of Information \& communications Technology Planning \& Evaluation (IITP) grant and National Research Foundation of Korea (NRF) grant funded by the Korea government (MSIT) (No. 2019-0-00421, 2022-0-00680-003, IITP-2024-2020-0-01821, and NRF-2018R1A5A1060031).

\bibliographystyle{ACM-Reference-Format}
\bibliography{references}

\end{document}